\documentclass[12pt]{article}
\usepackage{amssymb}
\usepackage{epsfig}

\catcode`\@=11
\def\numberbysection{\@addtoreset{equation}{section}
        \def\theequation{\thesection.\arabic{equation}}}
\def\half{\frac{1}{2}}
\def\beq{\begin{equation}}
\def\eeq{\end{equation}}
\numberbysection
\begin{document}
\begin{titlepage}
\begin{center}
\hfill DFF  1/7/99 \\
\vskip 1.in
{\Large \bf (2+1)-AdS Gravity on Riemann Surfaces}
\vskip 0.5in
P. Valtancoli
\\[.2in]
{\em Dipartimento di Fisica dell' Universita', Firenze \\
and INFN, Sezione di Firenze (Italy)}
\end{center}
\vskip .5in
\begin{abstract}
We discuss a formalism for solving $(2+1)$ $AdS$ gravity on Riemann
surfaces. In the torus case the equations of motion are solved by two
functions $f$ and $g$, solutions of two independent $O(2,1)$ sigma
models, which are distinct because their first integrals contain a
different time dependent phase factor. We then show that with the
gauge choice $k = \sqrt{\Lambda} / tg( 2\sqrt{\Lambda} t) $ the same
couple of first integrals indeed solves exactly the Einstein equations
for every Riemann surface. The $X^A = X^A(x^\mu)$ polydromic mapping
which extends the standard immersion of a constant curvature
three-dimensional surface in a flat four-dimensional space to the case
of external point sources or topology, is calculable with a simple
algebraic formula in terms only of the two sigma model solutions $f$
and $g$. A trivial time translation of this formalism allows us to
introduce a new method which is suitable to study the scattering of
black holes in $(2+1)$ $AdS$ gravity.
\end{abstract}
\medskip
\end{titlepage}
\pagenumbering{arabic}
\section{Introduction}

In this article we extend a previous study of $(2+1)$ gravity on
Riemann surface to the case of cosmological constant
\cite{a1},\cite{a2},\cite{a3},\cite{a4},\cite{a5},\cite{a6},\cite{a7}.
The absence of gravitational radiation introduces an important
simplification to the dynamics since it allows to choose an
instantaneous gauge for the propagation of the gravitational field,
and we can forget the problem of the delay due to the speed limitation
of signals.

The classical gravitational dynamics reduces to a renormalization of
the matter sources, and in particular the two-body interacting problem
can be solved as the two particles move following the geodesics around
an effective source, whose invariant mass is computable with the Wilson
loop of the spin connection around the two particles \cite{a7}.

The case of the cosmological constant addition is useful not only at a
classical level, where it sheds light on black-hole scattering, but
also at a quantum level because it teaches us how to quantize the
gravitational field when it doesn't carry bulk degrees of freedom
\cite{a4}. It
seems that although non-renormalizable $2+1$ gravity can be understood
at the quantum level by virtue of its integrability, of being a
topological theory, whose dynamics is to introduce a braiding
of a two-dimensional boundary theory, which can be related to a
conformal field theory or some integrable deformation of it.

At a quantum level we expect that the gravitational degrees of freedom
have lower dimension  and live on the boundary of the bulk. Several
articles have recently emphasized the importance of the holography in
field theory, as the property that allows to reconstruct the field in
the bulk starting from the knowledge of the field at the boundary.
In $2+1$ gravity the holography property seems particularly easy to show.

Our main aim will be investigate how to reduce classical $AdS$ gravity
to a two dimensional field theory. Let us remember rapidly the main
results obtained in this direction. At a classical level particle
dynamics has already been solved in the gauge $k=0$ \cite{a10} since
this gauge makes possible to reduce the interacting problem to a
choice of conformal mappings, defined by the monodromy conditions,
related to the particles' constant of motion. Exact results have been
found in the two-body case \cite{a10}, and an interesting connection
with Painlev\'e VI for the three body case \cite{a10,a12}. Furthermore
in order to treat topological degrees of freedom we followed Moncrief
choice of a time slicing  ( $k$ constant but not zero )\cite{a19}. The
resulting equations of motions can be simplified by choosing a
conformal gauge for the spatial metric. It has been recognized that
the field dynamics for Riemann surfaces can be reduced to a
sine-Gordon theory, at least for the conformal factor of the spatial
metric, and a first-order formalism based on a first-integral of a
$O(2,1)$ sigma model can be built. In this first-order formalism we
have built out of the solution $f$ of the $O(2,1)$ sigma model a
coordinate transformation $X^a = X^a(x^\mu)$ from a Minkowskian
coordinate system to the physical one.
     Each spatial slice is equivalent to a Riemann surface,
which we represent with branch cuts on the complex plane. The
corresponding metric is singular at the branch points, which move as
particle singularities. The motion of the branch points in the $X^A$
coordinates is free and determined by the Poincar\'e holonomies,
defining the coupling of Riemann surface to gravity. By solving the
mapping $f$, we can find the dynamics of the branch points and of the
moduli of the Riemann surface in the physical coordinate system
\cite{a8}.

At a level of the cosmological constant we are going to show that it
is possible to avoid further complications, and the first-order
formalism can be easily generalized to a couple of first-integrals of
two $O(2,1)$ sigma-models, distinct only by a phase factor in the
source. As a first consequence we can generalize the classical
mathematical theorem regarding the immersion of a Riemann surface with
genus $g>1$ in the Poincare' disk to the immersion of $AdS$ gravity on
Riemann surfaces into the direct product of two Poincare' disks.
Then the mathematical formalism which has been successfully applied to the
$(2+1)$-gravity case can be generalized without problem to the $AdS$
case.

A simple time translation of the gauge choice produces the gauge condition
suited to study the scattering of two particles and, a case more
interesting, the scattering of black holes. As an outcome of our work
we give at disposition an useful reference-point for the study of this
non-perturbative classical problem of general relativity,
apart from the fact that conformal gauge choice, which is peculiar to all our
work, can be done only in the ( connected ) region outside the horizons.

These are, from our point of view, the more appropriate gauge choices in
which to look for a solution for the Riemann surfaces coupled to $AdS$
gravity and the black hole scattering. While for the former ones we
give the explicit solution in some cases, for the second problem we
postpone to a future work the detailed study.

At a classical level, for example the torus, we recover that its
modulus describes a circular motion in Teichmuller space, while for
certain Riemann surfaces defined by the condition  $f=g$ the solutions
to the two sigma models become analytic and the moduli of the Riemann
surfaces are static, i.e. they do not have any temporal dependence.
Only with a non-analytic solution of the two $O(2,1)$ sigma models
it is possible to introduce a non trivial dynamics.

At a quantum level since all these theories have a finite number
of degrees of freedom we expect a reduction of quantum field theory to
a problem of quantum mechanics. For example we can quantize the
Lagrangian of the modulus of the torus, introducing a canonical
momentum to $\tau$, and writing down a Schr\"odinger equation acting
on the Hilbert space of square integrable functions of $\tau$
\cite{a8}. This reduced quantization is analogous to the particle case
\cite{a4}, where one can integrate out the field into an effective
action for the particle degrees of freedom. This reduced quantization
is an useful short-cut of the full second-quantized problem, i.e. in
which one ask himself how to treat topology-changing amplitudes or
creation and annihilation processes for particles \cite{a1}.

Maybe a new treatment of three dimensional gravity resembling
two dimensional quantum field theory can handle these problems.
In this sense the idea of 't Hooft of quantizing gravity directly in
the singular cordinate system seems particularly fruitful, because all
the classical results have this common characteristic. Moreover there
is in conformal field theory the example of Kniznik with his idea to
associate the fields of a conformal field theory to the sheets of a
Riemann surfaces. A further investigation in this direction is tempting.

\section{ York time gauge in the second order formalism }

We shall adopt in the following the ADM formalism, since all our
result take the assumption that space-time can be globally decomposed
as $\Sigma (t) \otimes R$, where $\Sigma(t)$ is a set of space-like
surfaces \cite{a13,a14}. A particularly meaningful parameterization
for the metric revealing the physically relevant degrees of freedom
turns out to be this one:

\beq ds^2 = \alpha^2 dt^2 - e^{2\phi} {| dz - \beta dt |}^2 ,
\label{2.1} \eeq
where we have chosen conformal coordinates for the spatial metric.
This choice of variable, the lapse function
$\alpha$ and the shift functions $\beta$, is particularly useful
when we discuss how to solve the Eulero-Lagrange equations of motion.

The $ADM$ decomposition can be performed at the level of the Einstein-Hilbert
action by splitting it into a spatial part,
intrinsic to the surfaces $\Sigma (t)$, and an
extrinsic part, coming from the embedding , as follows

\begin{eqnarray} S & =  &
- \half \int \sqrt{|g|} \ R^{(3)} \ d^3 x + \Lambda \int
\ \sqrt{|g|} d^3 x = \nonumber \\
& - & \half \int \sqrt{|g|} \ \left[
R^{(2)} + {( Tr K )}^2 - Tr ( K^2 ) \right] \ d^3 x + \Lambda \int
\ \sqrt{|g|} d^3 x
, \label{2.2} \end{eqnarray}

The extrinsic curvature tensor $K_{ij}$ , or second fundamental form
of the surface $\Sigma(t)$, is given in terms of the covariant derivatives
$\nabla^{(2)}_i$ with respect to the spatial part of the metric:

\beq K_{ij} = \half \sqrt{ \frac{|g_{ij}|}{|g|} } \left( \nabla^{(2)}_i g_{0j}
+ \nabla^{(2)}_j g_{0i} - \partial_0 g_{ij} \right). \label{2.3} \eeq

Our aim is to simplify at the maximum level the $ADM$ scheme applied
to $(2+1)$ $AdS$ gravity on Riemann surfaces . To achieve such
simplification a preliminary step is choosing the conformal
coordinates for the spatial metric, which is general if we allow to
represent a Riemann surface on a complex plane with branch points (
four branch points for a torus, $2g+2$ for an hyperelliptic surface of
genus $g$, and so on ). This means that the metric has a singular
particle-like behaviour on such branch points, which however must
satisfy some integrability condition, such as a finite area $ A(t) =
\int d z d {\overline z} e^{2 \phi} $ on each spatial slice.

The lagrangian for ($2+1$)-gravity, restricted to a spatial metric in
conformal gauge, is defined by:

\begin{eqnarray}
 {\cal L} & = & \alpha \nabla^2 \phi + \alpha e^{2\phi} K^2 -
\frac{e^{2\phi}}{\alpha} {| \partial_z \overline{\beta} |}^2 - \Lambda
\alpha e^{2\phi}
\nonumber \\
k & = & - \half g^{ij} K_{ij} =
\frac{e^{-2\phi}}{2\alpha} [ \partial_z ( \beta e^{2\phi} ) +
\partial_{\overline z} ( \overline \beta e^{2\phi} ) + \partial_0
e^{2\phi} ] \label{2.4} .\end{eqnarray}

A simplifying feature appears in the equation of motion for $\beta$,
which reads

\beq \partial_{\overline z} N + e^{2\phi} \partial_z K = 0 , \  \ \
N = \frac{e^{2\phi}}{2\alpha} \partial_z \overline \beta .
\label{2.5} \eeq

From this equation it is clear that the function $N$ is analytic
whenever $k=0$
or $k=k(t)$ is a time-dependent constant. In the following we will
show that a convenient choice is
\beq \ \ \ \ k = \frac{\sqrt{\Lambda}}{tg(2\sqrt{\Lambda}t)}
 \ \ \rightarrow  \ \ \ \ \ \partial_{\overline
z} N = 0 . \label{2.6} \eeq

Therefore our gauge choice is defined by the conditions

\beq g_{zz} = g_{{\overline z}{\overline z}} = 0 \ \ \ \ \ k =
\frac{\sqrt{\Lambda}}{tg(2\sqrt{\Lambda}t)} \label{2.7},  \eeq

and thus corresponds to a conformal gauge, with York time $g_{ij}
K^{ij} = - 2\frac{\sqrt{\Lambda}}{tg(2\sqrt{\Lambda}t)} $.

The above conditions are enough to eliminate time derivatives from the
Lagrangian and to give an instantaneous propagation of the metric,
as it appears from the Eulero-Lagrange equations for $\alpha$ and $\phi$:

\begin{eqnarray}
\nabla^2 \phi  +  4 N \overline{N} e^{-2\phi} & = & ( k^2 (t) +
\Lambda ) e^{2\phi} \nonumber
\\ \partial_{\overline{z}} N(z) & = & 0 \nonumber \\
\nabla^2 \alpha - 8 N \overline{N} e^{-2\phi} \alpha & = &
2 ( k^2 + \lambda ) \alpha e^{2\phi} + 2 \partial_0 k(t) e^{2\phi} = 2
( k^2 (t) + \Lambda )( \alpha - 2 ) e^{2\phi}. \label{2.8}
\end{eqnarray}

To write the last line we have used the property

\beq \partial_0 k(t) = - ( k^2(t) + \Lambda ), \label{2.9} \eeq

that works only if

\beq k(t) = \frac{ \sqrt{\Lambda}}{tg(2\sqrt{\Lambda}t)} \ \ \ \ \ \ { or
} \ \ \ \ k(t) = - \frac{ \sqrt{\Lambda}}{cotg(2\sqrt{\Lambda}t)}.
\label{2.10} \eeq

It turns out hat the first choice is useful to describe the solutions for
Riemann surfaces, while the second one is useful to write down
solutions for scattering of particles and black holes.

These equations are difficult to solve directly. However, we will show
that in the first-order formalism it naturally appears a quite simple
structure, related to a couple of  $O(2,1) \ $ $\sigma$-models,
which automatically solves them.

Although not explicit, the appearance of singularities in the metric
produce extra $\delta$ function sources, localized on the branch
points, in the equations of motion (\ref{2.8}). The $t$ dependence of the
Riemann surface moduli is therefore provided by the covariant
conservation of the ``underlying'' energy-momentum tensor, which in
turn implies the geodesic equations for the branch point singularities.

\section{ Some solutions in the first-order formalism }

The first-order formalism gives a direct language for relating
holonomies to the physical metric \cite{a15,a16}. For example, in \cite{a10}
we have proposed a non-perturbative solution for the metric and the
motion of $N$ interacting spinless particles in $(2+1)$ gravity,
based on a harmonic mapping $X^A = X^A (t, z, {\overline z})$ from
a regular coordinate system to Minkowskian multivalued coordinates.

Analogously to what we found for the gauge $k=0$, we are going to
solve the gauge condition which corresponds to the solution of the
torus coupled to $AdS$ $(2+1)$-gravity , given by

\beq k = k(t) = \frac{\sqrt{\Lambda}}{tg ( 2 \sqrt{\Lambda} t)}
\ \ \ \  g_{zz} = 0. \label{3.1} \eeq

The introduction of the cosmological constant makes useless searching
for a simplification of the equations of motion in the dreibein
formalism \cite{a33}, while it is fruiful to start from the typical
construction of space-times with constant curvature \cite{a31}, that are
obtained as an embedding in a four-dimensional flat metric with
signature (++--):

\beq ds^2 = dX^A dX^B \eta_{AB} \ \ \ \ \ \ \ \ X^A X_A =
\frac{1}{\Lambda} .\label{3.2} \eeq

This formalism can also be considered a first-order formalism,
defining a new dreibein $E^A_\mu = \partial_\mu X^A$, in which the
Lorentz index depends on four coordinates.

An intrinsic local frame of the four-dimensional space-time is given
by these  four four-vectors $X^A, \partial_z X^A,
\partial_{\overline{z}} X^A$ and $V^A$, which is defined as
\beq V^A = 2i \sqrt{\Lambda} e^{_2\phi} \epsilon^{ABCD} X_B \partial_z
X_C \partial_{\overline{z}} X_D \label{3.3} \eeq orthogonal to the
first three ones, and defined to have norm equal to unity. Instead
$\partial_z X^A$ and $\partial_{\overline{z}} X^A$ are two null
vectors having a non-vanishing scalar product between them. The gauge
choice made in the second order formalism corresponds to the following
set of equations for the relative immersion $X^A$:

\begin{eqnarray}
\partial_z \partial_{\overline{z}} X^A & = & \frac{\Lambda}{2} e^{2\phi}
X^A + \frac{k(t)}{2} e^{2\phi} V^A \nonumber \\
\partial^2_z X^A & = & 2 \partial_z \phi \partial_z X^A + N(z) V^A
\nonumber \\
\partial_z V^A & = & 2 e^{-2\phi} N(z) \partial_{\overline{z}} X^A +
k(t) \partial_z X^A . \label{3.4} \end{eqnarray}

We are going to introduce non trivial topology by allowing
multivaluedness of $X^A$. In fact the simplest
definition of a genus g Riemann surface coupled to $AdS$ gravity is the
quotient of the hyperboloid immersed in the $(2+2)$ flat space-time
by a finite set of
elements of the $SO(2,2)$ monodromy group, i.e. we identify all the
points that can be reached with isometries $(U_i, V_i , i=1,..,g)$ of
the flat metric $ds^2 = dX_0^2 + dX_1^2 - dX_3^2 - dX_4^2$ satisfying

\beq \prod_i U_i V_i U_i^{-1} V_i^{-1} = 1 . \label{3.5}\eeq
This means that circling many times around the cycles $ ( a_i, b_i, i
= 1, ...g )$ of the Riemann surface, the image of a point in the
regular coordinate system by the $X^A$ mapping is a lattice of points
in the hyperboloid immersed in a flat four-dimensional space-time.
The identification of this lattice with a point produces a Riemann
surface as spatial slice.

For the torus case, Eq. (\ref{3.5}) becomes $UV = VU$ and it is solved
by an abelian subgroup $Z\otimes Z$ of the $SO(2,2)$ monodromy group,
which can be taken as a boost along both the $X_0, X_3$ direction and
in the $X_1, X_2$ direction. The torus coupled to ($2+1$) $AdS$
gravity is defined by the following holonomy transformations:

\beq U : \left\{ \begin{array}{ccc}
X^0 & \rightarrow & ch \lambda_1 X^0 + sh \lambda_1 X^3 \\ X^3
&\rightarrow & ch \lambda_1 X^3 + sh \lambda_1 X^0 \\ X^1 &
\rightarrow & ch \lambda_2 X^1 + sh \lambda_2 X^2 \\ X^2 & \rightarrow &
ch \lambda_2 X^2 + sh \lambda_2 X^1 \end{array} \right.
\ \ \ V: \left\{
\begin{array}{ccc} X^0 & \rightarrow & ch \eta_1 X^0 + sh \eta_1
X^3 \\ X^3 &\rightarrow & ch \eta_1 X^3 + sh \eta_1 X^0 \\ X^1 &
\rightarrow & ch \eta_2 X^1 + sh \eta_2 X^2 \\ X^2 & \rightarrow &
ch \eta_2 X^2 + sh \eta_2 X^1 \end{array} \right. \label{3.6}
\eeq and the flat coordinates satisfy the usual quadratic relation

\beq {(X^0)}^2 + {(X^1)}^2 - {(X^2)}^2 - {(X^3)}^2 = \frac{1}{\Lambda}.
\label{3.7} \eeq

Based on the solution to the monodromies of the torus we have an
algebraic equation more than the standard condition of embedding:

\beq {X_0}^2 + {X_1}^2 - {X_2}^2 - {X_3}^2 = \frac{1}{\sqrt{\Lambda}} ,
\label{3.8} \eeq

which is given by

\beq \frac{ X_0^2 - X_3^2 }{ cos^2 ( \sqrt{\Lambda} t)} =
\frac{ X_1^2 - X_2^2 }{ sin^2 ( \sqrt{\Lambda} t)} . \label{3.9} \eeq
This condition can be interpreted as orthogonality condition between $X^A$
and $V^A$ :

\begin{eqnarray}
X^A & = &( X^0, X^1, X^2, X^3 ) \nonumber \\ V^A & = &
\sqrt{\Lambda}  \left[
- tg ( \sqrt{\Lambda } t ) X^0  , cotg ( \sqrt{\Lambda} t ) X^1 ,
cotg (\sqrt{\Lambda} t ) X^2, - tg ( \sqrt{\Lambda} t ) X^3
\right] . \label{3.10} \end{eqnarray}

If we apply many times from a generic point ($X^0, X^1, X^2, X^3$) the
transformations $(U, V)$ connected to the cycles $(a,b)$ of the torus,
we obtain a lattice of points which belong to the surface:

\beq X_0^2 - X_3^2 = \frac{\sqrt{1-C^2}}{\sqrt{\Lambda}} \ \ \ \
X^2_1 - X^2_2 = \frac{C}{\sqrt{\Lambda}}
. \label{3.11} \eeq

At a given $C$ such a surface describes again a torus, and the
space-time evolution is simply obtained by allowing a time-dependent
constant $C = C(t)$, which in the York time gauge is

\beq C(t)  = sin(\sqrt{\Lambda}t) . \label{3.12} \eeq

After a transformation of the flat coordinates

\begin{eqnarray}
 X^0 & = & \frac{cos(\sqrt{\Lambda} t)}{\sqrt{\Lambda}}
 \cosh \left( \frac{\sqrt{\Lambda} Y}{cos(\sqrt{\Lambda} t)} \right)
\ \ \ X^1 = \frac{sin(\sqrt{\Lambda} t)}{\sqrt{\Lambda}}
 \cosh \left( \frac{\sqrt{\Lambda} U}{
sin(\sqrt{\Lambda}t)} \right)
\nonumber \\ X^2 & = & \frac{sin(\sqrt{\Lambda} t)}{\sqrt{\Lambda}}
 \sinh \left( \frac{\sqrt{\Lambda} U}{sin(\sqrt{\Lambda} t)} \right)
\ \ \ X^3 = \frac{cos(\sqrt{\Lambda} t)}{\sqrt{\Lambda}}
 \sinh \left( \frac{\sqrt{\Lambda} Y}{
cos(\sqrt{\Lambda}t)} \right)
, \label{3.13} \end{eqnarray}
the lattice of points on the hyperboloid (\ref{3.8}) becomes a plane
lattice in
the $(U,Y)$ coordinates , which is
analogous to a static torus. The rescaling $t$-dependent factor  in front of
$U$ and $Y$ is necessary to keep the spatial metric in conformal gauge:

\begin{eqnarray}
ds^2 & = & \left[ 1 - \Lambda \left( \frac{ Y^2 sin^2 ( \sqrt{\Lambda} t
)}{ cos^2 ( \sqrt{\Lambda} t )} + \frac{ U^2 cos^2 ( \sqrt{\Lambda} t
)}{ sin^2 ( \sqrt{\Lambda} t )}\right) \right] dt^2
+ 2 \frac{\sqrt{\Lambda} cos ( \sqrt{\Lambda} t )}{sin ( \sqrt{\Lambda} t )}
U dU dt - \nonumber \\
& - &  2 \frac{\sqrt{\Lambda} sin ( \sqrt{\Lambda} t )}{cos ( \sqrt{\Lambda} t )}
Y dY dt - dU^2 - dY^2 , \label{3.14} \end{eqnarray}

from which we read the torus solution in the second order formalism
$\alpha = 1$, $\beta = \frac{\sqrt{\Lambda} cos
(\sqrt{\Lambda} t )}{sin (\sqrt{\Lambda} t )} U -i \frac{\sqrt{\Lambda} sin
(\sqrt{\Lambda} t )}{cos (\sqrt{\Lambda} t )} Y$  and $e^{2\phi}= 1$.

The set of holonomy transformations (\ref{3.6}) become the following
ones which are pure translation monodromies ( $\widetilde{Z} = U + i Y
$) :
\begin{eqnarray} {\widetilde Z}
& \stackrel{a}{\longrightarrow} &
{\widetilde Z} + \frac{1}{\sqrt{\Lambda}} [ \lambda_1
sin ( \sqrt{\Lambda} t ) + i \lambda_2 cos ( \sqrt{\Lambda} t ) ]
\nonumber \\
\ \ \ \ \   & \stackrel{b}{\longrightarrow} & {\widetilde{Z}} +
\frac{1}{\sqrt{\Lambda}} [ \eta_1 sin (
\sqrt{\Lambda} t ) + i \eta_2 cos ( \sqrt{\Lambda} t ) ]
. \label{3.15} \end{eqnarray}

The solution to them  can be represented as a standard abelian integral on a $z$-plane
with two branch cuts
\beq {\widetilde Z} = \int^z_0 \frac{dz}{w(z,t)} \ \ \ \ \ \ \ \  w^2 (z,t) = R(t)
z ( z - 1) ( z - \xi (t) ) , \label{3.16} \eeq
where the position of the third singularity $\xi(t)$ is time-dependent,
in order to allow that the translation monodromies are time-dependent
(\ref{3.15}).

Therefore, the solution for the torus is given by the composition of
the mapping (\ref{3.13}) with the abelian integral mapping (\ref{3.16}).

The  holonomies (\ref{3.6}) tell us that the motion of the branch
points is apparently almost free in the $X^A$ coordinates, as they
move as geodesics of the hyperboloid  (\ref{3.7}). Let us recall that
a generic  geodesic motion of it is parametrized by the following
equation:
\begin{eqnarray}
X^A & = & \frac{1}{\sqrt{\Lambda}} ( c^A_0 cos ( \sqrt{\Lambda} t ) +
c^A_1 sin ( \sqrt{\Lambda} t ) ) \nonumber \\
c^A_0 c_{A0} & = & 1 \ \ \ \ \ c^A_1 c_{A1} = m \ \ \ \ \  c^A_0 c_{A1} =
0 \label{3.17} \end{eqnarray}
where $m$ is equal to $1, 0, -1$ respectively for a timelike, null or
spacelike geodesic.

Taking for example the particle in $0$ at rest, the resulting values
for the constants $c^A_i, i = 1,2 $ are the following:

 \begin{eqnarray}
c^A_0 (0) & = & (1,0,0,0) \ \ \ \ c^A_1 (0) = (0,1,0,0)
\nonumber \\
c^A_0 (1) & = &
(cosh(\frac{\lambda_2}{2},0,0,sinh(\frac{\lambda_2}{2}))
\ \ \ \ c^A_1 (1) = (0,cosh(\frac{\lambda_1}{2},sinh(\frac{\lambda_1}{2}),0)
\nonumber \\
c^A_0 (\xi) & = &
(cosh(\frac{\eta_2}{2},0,0,sinh(\frac{\eta_2}{2}))
\ \ \ \ c^A_1 (1) = (0,cosh(\frac{\eta_1}{2},sinh(\frac{\eta_1}{2}),0).
\label{3.18} \end{eqnarray}
In the $z$-coordinates, only the motion of the third singularity
$\xi(t)$ is necessary up to a suitable rescalings of the $z$-coordinate,
while the other two can remain at rest.

It is straightforward to derive the equations of motion for the modulus
of the torus and for the area, which take the usual form \cite{a7},\cite{a8}:

\beq \tau (t) = \frac{ \lambda_1 sin(\sqrt{\Lambda}t) + i \lambda_2
cos(\sqrt{\Lambda}t) }{ \eta_1 sin(\sqrt{\Lambda}t) + i \eta_2
cos(\sqrt{\Lambda}t) }
\ \ \ \  \ \ A(t) \ = \ \int  dz d {\overline z} \ e^{2\phi} =
| \frac{sin(2\sqrt{\Lambda} t)}{2}   \ (\lambda_1 \eta_2 -
\lambda_2 \eta_1 ) | . \label{3.19} \eeq

The motion of the modulus is essentially a consequence of the free motion
in flat coordinates of the branch points. It has also to
satisfy another consistency condition \cite{a8}, namely that the motion of the
moduli must be geodesic with respect to the natural
metric of the moduli space, the Weil-Petersson metric, which in the
case of the torus is equivalent to the Poincar\'e metric of the upper
half $\tau$-plane:

\beq ds^2_{\tau} = \frac{d \tau d {\overline \tau } }{ {(Im  \tau )}^2 } .
\label{3.20} \eeq

The form (\ref{3.19}) of the solution for the modulus is consistent
since it describes a circular arc in the moduli space, which is a
geodesic of the Poincar\'e metric.

The corresponding dreibein $E^A_\mu = \partial_\mu X^A$ is given by
( A =  ( 0,1, x, y )) :

\beq E^A_z = \frac{\sqrt{\Lambda}}{2} \left( \frac{X^3}{i cos(\sqrt{\Lambda}
t)}, \frac{X^2}{ sin(\sqrt{\Lambda}t)},  \frac{X^1}{
sin(\sqrt{\Lambda}t)},
\frac{X^0}{i cos(\sqrt{\Lambda}t)} \right) . \label{3.21} \eeq

The conformal gauge condition ${(E^A_z)}^2 = 0 $ is verified due to
Eq. (\ref{3.9}).

Let us rewrite the torus mapping in terms of the general parametrization

\begin{eqnarray} \left( \begin{array}{cc} X^t & X^z \\ \overline{X}^z & \overline{X}^t
\end{array} \right) & = & \frac{1}{\sqrt{\Lambda}} {}^4
\sqrt{\frac{ \partial_z g \partial_{\overline{z}} \overline{f} }{ \partial_z f \partial_{\overline{z}}
\overline{g} } } \frac{e^{-i T}}{ \sqrt{(1-f\overline{f})(1-g\overline{g})} }
\left( \begin{array}{cc} - f \overline{g} & i f \\ i \overline{g} & 1
\end{array} \right) + \nonumber \\
& + & \frac{1}{\sqrt{\Lambda}} {}^4\sqrt{\frac{ \partial_z f
\partial_{\overline{z}} \overline{g} }{ \partial_z g \partial_{\overline{z}}
\overline{f} } }\frac{e^{i T}}{ \sqrt{(1-f\overline{f})(1-g\overline{g})} }
\left( \begin{array}{cc} 1 & -i g \\ -i \overline{f} & -\overline{f} g
\end{array} \right) . \label{3.22} \end{eqnarray}

It is enough to choose
\begin{eqnarray} f & = & th \left(
\frac{\sqrt{\Lambda}U}{2sin(\sqrt{\Lambda}t)} +
\frac{\sqrt{\Lambda}Y}{2cos(\sqrt{\Lambda}t)}
\right) \nonumber \\
g & = &  th \left(
\frac{\sqrt{\Lambda}U}{2sin(\sqrt{\Lambda}t)} -
\frac{\sqrt{\Lambda}Y}{2cos(\sqrt{\Lambda}t)}
\right) \label{3.23} , \end{eqnarray}

from which

\beq N(z) = \frac{\sqrt{\Lambda}}{2 sin ( 2 \sqrt{\Lambda} t ) w^2 }
\label{3.24} \eeq

and the whole solution reduces to a couple of first integrals of two
$O(2,1)$-$\sigma$ models:

\begin{eqnarray}
\frac{ \partial_z f \partial_z \overline{f} }{ {( 1- f \overline{f} )}^2 } =
\frac{\sqrt{\Lambda}}{2} \frac{e^{-2i \sqrt{\Lambda} t}}{
sin (2\sqrt{\Lambda} t )} N(z) \nonumber \\
\frac{ \partial_z g \partial_z \overline{g} }{ {( 1-  g\overline{g}
)}^2 } =\frac{\sqrt{\Lambda}}{2} \frac{e^{2i \sqrt{\Lambda} t}}{ sin
(2\sqrt{\Lambda} t )} N(z) .
\label{3.25} \end{eqnarray}

So we have found that in the torus case both
$f$ and $g$ are real and $N$ is related to the quadratic holomorphic
differential. In general, we can guess that $N(z,t)$ is a known source
for Eq. (\ref{3.25}), being a combination of the quadratic holomorphic
differentials of the Riemann surface.

The function $\tanh$ can also be expected since it diagonalizes the
monodromy conditions for $f$ around the cycles $C_i= (a,b)$ of the torus:

\beq f \stackrel{C_i}{\longrightarrow}
\frac{A_i f + B_i}{B_i f + A_i} \ \ \ \ \ A_i = \cosh
\frac{\chi_i}{2} \ \  B_i = \sinh \frac{\chi_i}{2} \ \ \ \ \ i = 1,2 .
\label{3.26} \eeq
The linear dependence of its argument from the abelian integrals
represents the change of sign $f \rightarrow -f $ around each branch point.

Since $f$ and $g$ are real in the case of the torus and their image are
contained inside the unit disk, they map a real
variable  inside the real diameter $D = [-1, 1]$.
 Since $f$ and $g$ are polydrome, they can be restricted to
cover a segment of $D$. Circling many times around each cycle of the
torus, $Image \ f$ and $Image \ g$ give a tessellation of the diameter.

This particular feature of the torus should be valid in general. For
every Riemann surface the image of the maps $f$ and $g$ can be restricted to
a polygon inside the corresponding unit disk.
Circling many times around each cycle of the Riemann surfaces
we should obtain a tessellation of the two unit disks, instead of their
diameters,
as a consequence of the nonabelian relation (\ref{3.5}) which replaces the
abelian one $UV = VU$ for the torus.

Another interesting case is the $\Lambda$-mapping for static Riemann surfaces,
which is  the case  $f = g, T = 2\sqrt{\Lambda} t$, where the
 $\Lambda$-mapping reduces to :
\begin{eqnarray}
X^A & = & \frac{1}{\sqrt{\Lambda}} \left[ cos ( 2 \sqrt{\Lambda} t ) ,
sin ( 2 \sqrt{\Lambda} t ) n^i \right] \nonumber \\ V^A & = & \left[ -
sin( 2 \sqrt{\Lambda} t ), cos ( 2 \sqrt{\Lambda} t ) n^i \right] ,
\nonumber \\
n^i & = & \left( \frac{1+f\overline{f}}{1-f\overline{f}} ,
\frac{2\overline{f}}{1-f\overline{f}} , \frac{2f}{1-f\overline{f}} \right).
\label{3.28} \end{eqnarray}

A similar mapping has been discussed in \cite{a35}. The particular
relation (\ref{3.28}) between flat $X^A$-coordinates and $f$ realizes
explicitly the isomorphism between the subgroup $SO(2,1)$ of $SO(2,2)$
and $SU(1,1)$, and the M\"obius transformations of $f$ correspond to
the $SO(2,1)$ holonomies for the $X^A$ coordinates.

We easily recognize that the flat coordinates satisfy the
constraint:

\beq {(X^1)}^2 - X^z X^{\overline{z}} = \frac{sin^2
( 2\sqrt{\Lambda} t ) }{\Lambda}. \label{3.30} \eeq

The same surface can be obtain starting from a generic point
$X^A_0$ and  applying all the $SO(2,1)$
movements related to the cycles $(a_i , b_i)$.  Hence this time
foliation is natural since it is induced by the holonomies.

The $X= X(x)$ mapping (\ref{3.28}) produces the hyperbolic metric on the disk:

\beq ds^2 = 4 dt^2 - 4 \frac{sin^2 T}{\Lambda} \frac{ df d{\overline f}}{{(1 -
{|f|}^2)}^2} , \label{3.31} \eeq
from which we can read $\alpha = 2$, $\beta = 0 $ and $e^{2\phi} =
4 \frac{sin^2 T}{\Lambda} / {{( 1 - {|f|}^2 )}^2}$ which solve eq. (\ref{2.8}).

The conformal mapping $f(\overline z)$ still has to be determined.
Firstly, we remember that every genus g Riemann surfaces is determined by
a fundamental group $\pi_1 ( \Sigma )$ generated by the $2g$ holonomies
( $U_i , V_i$) satisfying the relation (\ref{3.5}). Let us denote with $H$
the unit disk with its metric of negative constant curvature. The group
$SU(1,1)$ acts on it, maintaining its metric. Consider a subgroup of it $\Gamma
\subset SU(1,1)$, isomorphic to $\pi_1 ( \Sigma )$, the quotient $ H / \Gamma$
is a Riemann surface of genus $g$, with the same constant curvature
metric of the unit disk $H$.

Therefore, we can think that the image of the conformal mapping $f$
gives a tessellation of the unit disk on which the holonomy acts as
$SU(1,1)$ and we can restrict the fundamental region of $Im f$ to
be inside a closed geodesic $4g$-gon of the unit disk with hyperbolic metric,
where the conformal mapping becomes one to one.

For pure $SO(2,1)$ holonomies, the equations of motion for the branch points
are simply trivial timelike geodesic of the constraint $X^A X_A = \frac{1}{\Lambda}$.

Since the $X^A = X^A (x)$ mapping in Eq. (\ref{3.28}) has also a
trivial time dependence, we
conclude that there is no evolution in the $z$-coordinate for the
branch points. As a consequence, there is no time evolution for the
moduli \cite{a8}, and the only time dependence comes from the overall scale
factor $sin^2 T$ in Eq. (\ref{3.31}). We can
suppose that the branch points $\xi_i$ have fixed positions on the real axis.
Then to find the Gauss map $f$ it is helpful this theorem of complex
analysis: given a simple and closed geodesic polygon $\Pi$ of the Poincar\'e
metric in the upper half $w$-plane, whose sides are circular arcs making
angles $\alpha_1 \pi$, $\alpha_2 \pi$, ..., $\alpha_n \pi$, at the
vertices $A_1, A_2, ..., A_n$ where $ 0 \le \alpha_j \le 2 $, then
there exist real numbers $\xi_1, \xi_2, ... \xi_n$, $ \beta_1 , \beta_2 ,
... \beta_n$ such that

\begin{eqnarray} \xi_1 < \xi_2 < ... < \xi_n , \ \ \ \ \ \
\sum_{j=1}^{n} \beta_j = 0 \nonumber \\
\sum_{j=1}^{n} ( 2 \beta_j \xi_j + 1 - \alpha^2_j ) = 0 , \ \ \ \
   \sum_{j=1}^{n} ( \beta_j {\xi}^2_j + ( 1 - \alpha^2_j ) \xi_j ) = 0
\label{3.32} \end{eqnarray}
and the upper $z$-plane is conformally mapped inside $\Pi$ by

\beq w(z) = \frac{u_1 (z)}{u_2 (z)} , \label{3.33} \eeq
where $u_1 (z)$ and $u_2 (z)$ are two linearly independent solutions
of the Fuchsian differential equation:

\beq u'' + \left[ \frac{1}{4} \sum^{n}_{j=1}
\frac{1-\alpha^2_j}{{(z-\xi_j)}^2} + \half \sum^{n}_{j=1} \frac{
\beta_j}{z-\xi_j} \right] u = 0 . \label{3.34} \eeq

The points $\xi_j$ correspond to the vertices $A_j$. A simple
conformal mapping relates the upper $w$-plane endowed
with the Poincar\'e metric to the $f$-unit disk with the hyperbolic metric.

Let us remark that in order to obtain the tessellation property of the
unit disk, the angles $\alpha_i \pi$ must be chosen as $\pi(1 - 1/ n_i)$, with
$n_i$ positive integers. Such a mapping problem has been investigated in
connection to Fuchsian groups and functions. In fact, to
each geodesic polygon of the $f$-unit disk, whose angles in the
vertices have the measure $ \pi(1- 1/ n_i)$, with $n_i$  positive integers, it
is connected a discrete group of movements $\Gamma$ and an analytic
function $z(f)$ defined in the $f$-unit disk, which is invariant under
the $\Gamma$ action on the $f$ variable.

\beq z \left( \frac{a_k f + b_k }{{\overline b}_k f + {\overline a}_k }
\right) =  z(f)   \ \ \ \ \ \ \Longleftrightarrow
\ \ \ \ \ \ \ \ z( X^A ) = z( \Lambda^{(k) A}_B X^B )
\ \ \ \ \ \forall k . \label{3.35} \eeq

The $\Gamma$ group of SU(1,1)
M\"obius transformations on the disk is called Fuchsian group.
The function $z(f)$ is a
Fuchsian function with respect to the group $\Gamma$. Instead the
inverse $f = f(z)$ is polydrome, and it can be restricted to map
the $z$-plane into the geodesic
polygon. Therefore we conclude that our conformal mapping is the inverse
of such fuchsian function $z(f)$. Examples of them can be built from
the Poincar\'e series, having a simple covariant transformation
under the Fuchsian group, from which we can obtain a Fuchsian
function,  invariant under the action of the Fuchsian group.

\section{General solution for the immersion equation}

The example of the torus can teach us a lot of information for the
general case. We are going to discuss that we have already found the
more general equations that constrain the dynamics of every Riemann
surface.

Let us introduce a set of notations to be more comprehensible.  In
general the $O(2,2)$ cuts can be decomposed as products of $SU(1,1)$
cuts as
\beq \left( \begin{array}{cc} X^t & X^z \\ \overline{X}^z & \overline{X}^t
\end{array} \right ) \rightarrow
\left( \begin{array}{cc} A_1 & B_1 \\ \overline{B}_1 & \overline{A}_1
\end{array} \right )
\left( \begin{array}{cc} X^t & X^z \\ \overline{X}^z & \overline{X}^t
\end{array} \right )
\left( \begin{array}{cc} A_2 & B_2 \\ \overline{B}_2 & \overline{A}_2
\end{array} \right ) . \label{4.0} \eeq
To make the monodromies more explicit we solved in \cite{a20} the
constraint $X^A X_A = \frac{1}{\Lambda}$ with a parametrization that
carries esplicit projective representation of each $SU(1,1)$:
\beq f \rightarrow \frac{A_1 f + B_1}{\overline{B}_1 f +
\overline{A}_1 } \ \ \ \
g \rightarrow \frac{A_2 g + B_2}{\overline{B}_2 g +
\overline{A}_2 } . \label{4.01}
\eeq
It turns out that the following choice is relative simple and general:
\begin{eqnarray}
X^A  & = & \frac{1}{\sqrt{\Lambda}} ( h\tilde{W}^A + h^{-1} W^A )
= \frac{1}{\sqrt{\Lambda}} ( \overline{h^{-1}}
\tilde{U}^A + \overline{h} U^A ) \ \ \ \ \ \ \
h = e^{\frac{\phi_g-\phi_f}{2} - i T } \ \
\nonumber \\ V^A_0 & = & i ( - h \tilde{W}^A + h^{-1} W^A )
= i ( - \overline{h^{-1}}
\tilde{U}^A + \overline{h} U^A ) \ \ \ \ \ \ \ \ \ \label{4.1}
\end{eqnarray}
where we define the following vectorial functions of $f$ and $g$, which
have not real components:
\begin{eqnarray}
W^A & = & \sqrt{\frac{\partial_{\overline{z}} g }{
\partial_{\overline{z}} f } } \frac{ (1, - \overline{g} f, -i \overline{g},
- i f ) }{ 1- g \overline{g} } \ \ \ \ \ \ \
\tilde{W}^A  = \sqrt{\frac{\partial_{\overline{z}} f }{
\partial_{\overline{z}} g }  } \frac{ (-\overline{f} g, 1,
 i\overline{f}, i g)}{ 1- f\overline{f} }  \nonumber \\
U^A & = & \sqrt{\frac{\partial_z \overline{f} }{
\partial_z \overline{g} } } \frac{ (1, - \overline{g} f, -i \overline{g},
- i f)}{ 1- f \overline{f} } \ \ \ \ \ \ \
\tilde{U}^A  = \sqrt{\frac{\partial_z \overline{g} }{
\partial_z \overline{f} } } \frac{ (-\overline{f} g, 1,
 i\overline{f}, i g)}{ 1- g\overline{g} } ,\label{4.2}
\end{eqnarray}
but the vectors $X^A$ and $V^A_0$ have real components, being
coordinates. This global vector covariance can be useful in
characterizing the scalar product of these vectors and their
derivatives in terms of the invariants under the global SO(2,2)
monodromy group.

As an example we can list the following identities :
\begin{eqnarray}
  W \cdot \tilde{W} & = & U \cdot \tilde{U} = 1/2 \nonumber \\
\tilde{W} \cdot \partial_z W & = & \frac{1}{2} ( \overline{H}_g -
\overline{H}_f  ) \ \ \ \  h^{-1} \partial_z h = \frac{1}{2}
\partial_z ( \phi_g - \phi_f) - i\partial_z T \nonumber \\
\tilde{W} \cdot U & = &
{( W \cdot \tilde{U} )}^{-1} = \frac{e^{\phi_f-\phi_g}}{2} .
\label{4.3}
\end{eqnarray}

We also introduce the definition of the various invariants that will
be useful in the following :
\begin{eqnarray}
I_f & = & \frac{\partial_z f \partial_z \overline{f}}{1-f\overline{f}}
\nonumber \\ e^{2\phi_f} & = & \frac{\partial_z
\overline{f} \partial_{\overline{z}} f}{1-f\overline{f}} \  \ \ \ \ \
\ e^{2\Omega_f} =  \frac{\partial_z f \partial_{\overline{z}}
\overline{f}}{1-f\overline{f}}  \nonumber \\
H_f & = & \frac{1}{2} \frac{\partial_z \partial_{\overline{z}} f }{
\partial_z f} + \frac{\overline{f} \partial_{\overline{z}} f
}{1-f\overline{f}} \ \ \ \
\ \ \ \overline{H}_f = \frac{1}{2} \frac{\partial_z
\partial_{\overline{z}} f }{
\partial_{\overline{z}} f} + \frac{\overline{f} \partial_z f
}{1-f\overline{f}} \nonumber \\
H_{\overline{f}} & = & \overline{\overline{H}_f} \ \ \
\overline{H}_{\overline{f}} = \overline{H_f} , \label{4.4}
\end{eqnarray}
and analogously for $g$.

Now let ask ourself the following question: can $V^A_0$  be identified
with the vector $V^a$ which appears in the immersion equations
(\ref{3.4}) ? The answer is negative and positive at the same time,
negative if we decide to work with general unconstrained fields,
positive in practice, since if we look for a simple gauge fixed
solution then it is possible to make such identification. Let us work
out the difference:

\beq V^A = \rho ( V^A_0 + \overline{\gamma} \partial_z X^A + \gamma
\partial_{\overline{z}} X^A) \label{4.5} \eeq

in which the coefficients $\gamma$ and $\rho$ are derived in such a way to
satisfy the properties of $V^A$:

\begin{eqnarray} \gamma & = & 2 e^{-2\phi} V^A_0 \cdot \partial_z X^A =
\frac{2 i}{\sqrt{\Lambda}} e^{-2\phi} [  \partial_z ln (h) -
\overline{H}_g + \overline{H}_f ] \nonumber \\
\rho^2 & = & \frac{1}{1 + \gamma \overline{\gamma} e^{2\phi}}. \label{4.6}
\end{eqnarray}
Moreover, it will useful in the following to compute the derivative of
this vector $\partial_z V^A_0$, that can be  again parametrized  in
terms of the basis of the 4 four-vectors

\beq \partial_z V^A_0 = \gamma_0 X^A + \gamma_1 V^A + \gamma_2 \partial_z X^A +
\gamma_3 \partial_{\overline{z}} X^A \label{4.7} \eeq

where

\begin{eqnarray}
\gamma_0 & = & - \frac{\gamma}{2}  e^{2\phi} \nonumber \\
\gamma_1 & = & - \rho^{-2} \partial_z \rho - \overline{\gamma} N(z) -
                  \frac{k}{2} \gamma e^{2\phi} \nonumber \\
\gamma_2 & = & k \rho^{-1} - e^{-2\phi} \partial_z ( \overline{\gamma}
               e^{2\phi} )\equiv \widetilde{k} \nonumber \\
\gamma_3 & = & 2 \rho^{-1} e^{-2\phi} N(z) - \partial_z \gamma
\equiv 2 \widetilde{N} e^{-2\phi} .\label{4.8}
\end{eqnarray}

The immersion equations (\ref{3.4}) are covariant, and their
information can be
encoded in corresponding scalar products,

\begin{eqnarray}
{(\partial_z X^A )}^2 & = & 0 \ \ \ \ \ \ \ \partial_z V^A \cdot
\partial_z X^A = - N(z) \ \ \ \ \ \ \ {(\partial_z V^A)}^2
= -2 k N(z) \nonumber \\
\partial_z X^A \cdot \partial_{\overline{z}} X^A & = & - \frac{1}{2} e^{2\phi}
\ \ \ \ \ \ \ \partial_z V^A \cdot \partial_{\overline{z}} X^A = -
\frac{k(t)}{2} e^{2\phi}
\ \ \ \ \  \ \ \partial_z V^A \cdot \partial_{\overline{z}} V^A = -
\frac{{k(t)}^2}{2} e^{2\phi} - 2 N \overline{N} e^{-2\phi} .\nonumber \\  & &
\label{4.10} \end{eqnarray}

Let us develop such requirements in two separate steps, the first one
is using only the first line of eq. (\ref{4.10})

\beq {(\partial_z X^A )}^2 = 0 \ \ \ \ \ \ \partial_z V^A_0 \cdot
\partial_z X^A = - \tilde{N}(z) \ \ \ \ \ \ \ {(\partial_z V^A_0 )}^2
= -2 \tilde{k} \tilde{N}(z) + \frac{\gamma^2_0}{\Lambda} + \gamma^2_1
\label{4.11} \eeq

where we have decided to dress the basic fields $N(z)$ and the constant
$k$ with terms coming from $\gamma$:

\begin{eqnarray}
\tilde{N} & = & \rho^{-1} N(z) - \frac{1}{2} e^{2\phi} \partial_z \gamma
\nonumber \\
\tilde{k} & = &  k \rho^{-1} - e^{-2\phi} \partial_z ( \overline{\gamma}
 e^{2\phi} ) . \label{4.12} \end{eqnarray}

Inserting the general parametrization (\ref{4.1}) in the first series of
identities (\ref{4.11}) we get the following constraints:

\begin{eqnarray}
 \partial_z W \cdot \partial_z \tilde{W} & = &  \frac{1}{2} \left[
{( \partial_z ln(h) )}^2 - 2 \partial_z ln (h) (
\overline{H}_g - \overline{H}_f  ) -
 \widetilde{N}(z) \widetilde{k} \right] +\frac{1}{4\Lambda} \gamma^2_0
+ \frac{\gamma^2_1}{4}
\nonumber \\
 {( \partial_z W )}^2 & = & h^2 \left[ \frac{\widetilde{k} + i
\sqrt{\Lambda}}{2} \widetilde{N}(z) - \frac{1}{4\Lambda} \gamma^2_0
- \frac{\gamma^2_1}{4} \right]
\nonumber \\
 {( \partial_z \widetilde{W} )}^2 & = & h^{-2} \left[ \frac{\widetilde{k}
- i \sqrt{\Lambda}}{2} \widetilde{N}(z) - \frac{1}{4\Lambda} \gamma^2_0
- \frac{\gamma^2_1}{4} \right] .\label{4.13}
\end{eqnarray}

On the other hand the explicit computation of these scalar product
based on their definition gives the following identities

\begin{eqnarray}
\partial_z W \cdot \partial_z \tilde{W} & = & - \frac{1}{2} {(
\overline{H}_f - \overline{H}_g )}^2
- \frac{1}{2} \left( I_f + I_g \right) \nonumber \\
{(\partial_z \tilde{W} )}^2 & = & \frac{ \partial_z g }{
\partial_{\overline{z}} g } e^{2\phi_f} = h^{-4} e^{-4i T } I_g
\nonumber \\
{(\partial_z W )}^2 & = &  \frac{ \partial_z f }{
\partial_{\overline{z}} f } e^{2\phi_g} = h^4 e^{ 4 i T } I_f .
\label{4.15} \end{eqnarray}

Putting all the information together we finally obtain the following
fundamental relations:

\begin{eqnarray}
{ ( \partial_z ln (h) + \overline{H}_f -
\overline{H}_g )}^2 & = & I_f ( h^2 e^{4i T} - 1 ) + I_g ( h^{-2}
e^{-4iT} -1 ) \nonumber \\
\widetilde{N} & = & \frac{ h^2 e^{4iT} I_f - h^{-2} e^{-4iT} I_g }{i
\sqrt{\Lambda}} \nonumber \\
\frac{\gamma^2_0}{4\Lambda} + \frac{\gamma^2_1}{4} & = &
\frac{\widetilde{k} - i \sqrt{\Lambda} }{2i \sqrt{\Lambda}} h^2 e^{4iT} I_f
-\frac{\widetilde{k} + i \sqrt{\Lambda} }{2i \sqrt{\Lambda}} h^{-2} e^{-4iT}
I_g . \label{4.16} \end{eqnarray}

In the torus case these relations are solved by

\beq I_f = \frac{\sqrt{\Lambda}}{2} \frac{e^{-i T}}{sin T} N(z) \ \ \
\  I_g = \frac{\sqrt{\Lambda}}{2} \frac{e^{i T}}{sin T} N(z) .
\label{4.17} \eeq
It seems not possible
to think that these equations satisfy with only one more
relation all the immersion identities, but in reality it is like this, and
the purpose of computing the second series of identities is to show
that eqs. (\ref{4.17}) are completely self-consistent.

If one choose the case  $f=g$ these equations reduce to the following ones

\begin{eqnarray}
{(\partial_z T )}^2 & = & - I_f {( e^{iT} - e^{-iT} )}^2 \nonumber \\
\widetilde{k} & = &  - \sqrt{\Lambda}\left[ \frac{1}{tg(2T)}
+ \frac{\frac{\gamma^2_0}{\Lambda}+ \gamma^2_1}{4 I_f sin(2T)} \right]
\nonumber \\
I_f & = &  \frac{ \sqrt{\Lambda} \widetilde{N} }{2sin(2T)} .
\label{4.18} \end{eqnarray}
This case is a little misleading, since in the point of view of the
present article it only corresponds to $N(z)=0$ and
$I_f= I_g=0$, in the final simplified gauge choice. However
one can recover from it another result, stated in
reference \cite{a21}, valid for a different gauge choice $k=0$, where such
simplification is not possible.

Let us compute the following scalar products

\begin{eqnarray}
i) A & = & \partial_z W^A \cdot \partial_{\overline{z}} U^A \nonumber \\
ii) B & = & \partial_z \tilde{W}^A \cdot \partial_{\overline{z}} U^A
\nonumber \\
iii) C & = & \partial_z W^A \cdot \partial_{\overline{z}} \tilde{U}^A
\nonumber \\
iv) D & = & \partial_z \tilde{W}^A \cdot \partial_{\overline{z}} \tilde{U}^A
\label{4.19}
\end{eqnarray}

by firstly solving the following system

\begin{eqnarray}
e^{2i T} A & + & e^{-2i T} D + e^{\phi_g-\phi_f} B + e^{\phi_f-\phi_g}
C + \partial_z ln(h) ( H_{\overline{f}} -H_{\overline{g}} ) + \nonumber \\
 & + & \partial_{\overline{z}} ln(k) ( \overline{H}_g -
\overline{H}_f  ) - \partial_z ln (h)
\partial_{\overline{z}} ln (k) =  - \frac{\Lambda}{2} e^{2\phi}
\nonumber \\
e^{2i T} A & - & e^{-2i T} D + e^{\phi_g-\phi_f} B - e^{\phi_f-\phi_g} C
 =  i \sqrt{\Lambda} \frac{\gamma_2}{2} e^{2\phi} \nonumber \\
e^{2i T} A & - & e^{-2i T} D - e^{\phi_g-\phi_f} B + e^{\phi_f-\phi_g} C
=  i \sqrt{\Lambda} \frac{\overline{\gamma}_2}{2} e^{2\phi}
\nonumber
\\ - e^{2i T} A & - & e^{-2i T} D + e^{\phi_g-\phi_f} B +
e^{\phi_f-\phi_g} C + \partial_z ln(h) ( H_{\overline{f}}
-H_{\overline{g}} ) + \nonumber \\
 & + & \partial_{\overline{z}} ln(k) ( \overline{H}_g -
\overline{H}_f  ) - \partial_z ln (h)
\partial_{\overline{z}} ln (k) = \frac{1}{\Lambda} \gamma_0
\overline{\gamma}_0 + \gamma_1 \overline{\gamma}_1
- ( \gamma_2 \overline{\gamma}_2 + \gamma_3 \overline{\gamma}_3 ) e^{2\phi}
\nonumber \\ & & \label{4.20} \end{eqnarray}

coming from the immersion equations (\ref{3.4})

\begin{eqnarray} \partial_z X^A \cdot \partial_{\overline{z}} X^A & = &
 - \frac{1}{2}  e^{2\phi} = \alpha_1
\ \ \ \ \ \ \partial_z V^A_0 \cdot \partial_{\overline{z}} X^A = -
\frac{\gamma_2}{2} e^{2\phi} = \beta_1 \nonumber \\
\partial_{\overline{z}} V^A_0 \cdot
\partial_z X^A & = & - \frac{\overline{\gamma}_2}{2} e^{2\phi} =
\beta_2 \ \ \ \ \ \
\partial_z V^A_0 \cdot \partial_{\overline{z}} V^A_0 =
\frac{1}{\Lambda} \gamma_0 \overline{\gamma}_0 + \gamma_1 \overline{\gamma}_1
- \frac{1}{2}
( \gamma_2 \overline{\gamma}_2 + \gamma_3 \overline{\gamma}_3 ) e^{2\phi}
= \alpha_2 .\nonumber \\ & & \label{4.21}
\end{eqnarray}

It is not difficult to compute their explicit expression based on their
definition (\ref{4.1})
\begin{eqnarray}
A & = & \frac{1}{2} e^{\phi_f+\phi_g} \left[ 1+ \frac{\partial_z
f \partial_{\overline{z}} \overline{g} }{\partial_{\overline{z}} f
\partial_z \overline{g} } \right] \nonumber \\
B & = & \frac{1}{2} e^{\phi_f - \phi_g} \left[ ( H_{\overline{f}} -
H_{\overline{g}} ) (
\overline{H}_f - \overline{H}_g ) - e^{{2\phi}_f}
- e^{{2\Omega}_g} \right] \nonumber \\
C & = & \frac{1}{2} e^{\phi_g - \phi_f} \left[ ( H_{\overline{f}} -
H_{\overline{g}} ) (
\overline{H}_f - \overline{H}_g )
- e^{{2\Omega}_f} - e^{{2\phi}_g} \right] \nonumber \\
D & = & \frac{1}{2}e^{\phi_f+\phi_g} \left[ 1+
\frac{\partial_{\overline{z}} \overline{f} \partial_z g }{
\partial_z \overline{f} \partial_{\overline z} g } \right] . \label{4.22}
\end{eqnarray}

By comparing these equations with the result coming from the immersion
we obtain finally :

\begin{eqnarray}
( \partial_z ln h + \overline{H}_f - \overline{H}_g ) (
\partial_{\overline{z}} ln k + H_{\overline{g}} - H_{\overline{f}} ) &
= &
- e^{2\phi_f} - e^{2\Omega_g} - \frac{\Lambda\alpha_1+\alpha_2}{2} + i
\sqrt{\Lambda} \frac{\beta_1-\beta_2}{2} = \nonumber \\
& = & - e^{2\phi_g} - e^{2\Omega_f} -
\frac{\Lambda\alpha_1+\alpha_2}{2} - i \sqrt{\Lambda} \frac{\beta_1-\beta_2}{2}
\nonumber \\
e^{\phi_f+\phi_g} \left[ 1 + \frac{\partial_z
f}{\partial_{\overline{z}} f }
\frac{\partial_{\overline{z}} \overline{g}}{\partial_z \overline{g} }
\right] & = & \frac{e^{-2iT}}{2} \left[
\Lambda \alpha_1 - \alpha_2 - i \sqrt{\Lambda} ( \beta_1 + \beta_2 )
\right] \nonumber \\
e^{\phi_f+\phi_g} \left[ 1 + \frac{\partial_{\overline{z}}
\overline{f}}{\partial_z \overline{f} }
\frac{\partial_z g }{\partial_{\overline{z}} g }
\right] & = & \frac{e^{2iT}}{2} \left[
\Lambda \alpha_1 - \alpha_2 + i \sqrt{\Lambda} ( \beta_1 + \beta_2 )
\right] . \label{4.28} \end{eqnarray}

Now let us discuss these identities in detail. For example if we make
 the assumption that  $f = real$ and $g=real$ as in the
torus case, we obtain the following identities $A=\overline{D}$ and
$B=C$. In particular if
we substitute directly the torus solution we get:

\begin{eqnarray} A & = & \frac{\Lambda cos T e^{-iT} }{4 w \overline{w}
sin^2 T} \nonumber \\ B & = & C = - \frac{\Lambda}{4 w\overline{w}
sin^2 T}
\nonumber \\ D& = & \overline{A} =  \frac{\Lambda cos T e^{iT} }{4 w
\overline{w} sin^2 T} . \label{4.24} \end{eqnarray}

In the general case $f=g$  by definition we get

\beq  A = - B = - C = D , \label{4.25} \eeq

while on the other hand we obtain that

\begin{eqnarray}
A & = &  \frac{e^{-2i T}}{4} \left[ \Lambda \alpha_1 - \alpha_2 - i
\sqrt{\Lambda} ( \beta_1 + \beta_2 ) \right]
\nonumber \\
B & = & \frac{e^{\phi_f-\phi_g}}{2} \left[ \partial_z ln h
\partial_{\overline{z}} ln k + \frac{\Lambda \alpha_1 + \alpha_2}{2} - i
\sqrt{\Lambda} \frac{\beta_1-\beta_2}{2} \right]
\nonumber \\
C & = & \frac{1}{2} \left[ \partial_z ln h
\partial_{\overline{z}} ln k + \frac{\Lambda \alpha_1 + \alpha_2}{2} + i
\sqrt{\Lambda} \frac{\beta_1-\beta_2}{2} \right] \nonumber \\
D & = &  \frac{e^{2i T}}{4} \left[ \Lambda \alpha_1 - \alpha_2 + i
\sqrt{\Lambda} ( \beta_1 + \beta_2 )
\right] . \label{4.26}
\end{eqnarray}

The compatibility of these two equations ( still for $f=g$) implies that

\begin{eqnarray}
Im ( \gamma_2) & = & 0 \nonumber \\
\sqrt{\Lambda} \gamma_2 & = & tg(2T) \left[ ( \frac{\gamma_2
\overline{\gamma}_2 + \gamma_3 \overline{\gamma}_3}{2} - \frac{\Lambda}{2}
) - ( \frac{\gamma_0 \overline{\gamma}_0}{\Lambda}
 + \gamma_1 \overline{\gamma}_1 )
e^{-2\phi} )  \right] \nonumber \\
\partial_z ln(h) \partial_{\overline{z}} ln(h) & = &
-\frac{e^{2\phi}}{4} \sqrt{\Lambda} \left[ \sqrt{\Lambda} - tg T
\gamma_2 \right] .\label{4.27}
\end{eqnarray}

Finally the torus case and the static Riemann surface case have in
common the following simple solution of all these constraints, namely
that $T=T(t)$ and  $\gamma = 0 $, and by consequence that $\gamma_0 =
\gamma_1 = 0 $ ,  $\gamma_2 = k $ e $\gamma_3 = 2 e^{-2\phi} N(z)$,
and moreover that $\widetilde{N} = N, \widetilde{k} = k$.

\section{The gauge $\gamma = 0 $}

In $2+1$ gravity the gauge conditions $k = 0$ and $g_{zz}=0$ were
enough to fix completely the fields. Here the same remark cannot be
applied. The reason for this strange behaviour of $AdS$ gravity lies
in the fact that are no natural boundary conditions on the fields at
infinity, therefore we  have more freedom to fix the gauge.

We are going to choose the boundary conditions in such a way that the
equations of motion can be expressed by the simplest choice. It is not
difficult to find such a gauge. In fact, we can be inspired by the
simplest solutions that we have at disposition, the solutions for the
torus and for the static Riemann surfaces, which have in common the
auxiliary condition $\gamma = 0 $. If we take seriously such a
condition in general, for every Riemann surface, we obtain a drastic
simplification in the expression for the equations of motion,
practically the simplest one that we have could written obeying all
the physical requirements we have discussed so far:

\beq I_f e^{iT} = I_g e^{-iT} = \frac{ \sqrt{\Lambda} N(z)}{2 sin T}
\ \ \ \ H_f = H_g = 0 . \label{5.1}\eeq

Moreover it is true that $\gamma_0 = \gamma_1 = 0, \  \widetilde{k} = k, \
\widetilde{N} = N $ and $ T= 2 \sqrt{\Lambda} t , \ \ k =
\frac{\sqrt{\Lambda}}{tg (T) }$.

We know that an analitic solution of the monodromy conditions for $f$
and $g$ is pratically impossible for particles, because it gives rise
to many problems, for example the fact that the horizon defined by the
condition $|f|=1$ doesn't match in generally with the condition
$|g|=1$, which is  a quite important constraint, to know where the
spatial slice of the universe ends, or the problem how to give motion
to the particles, since for an analitic solution the geodesic
equations are undefined ( see \cite{a20} for an analysis of such
problems ).
 For reasons of symmetry with the particle case we believe that  analogous
problems would also appear for Riemann surfaces.

This is the simplest equation for a non-analytic solution to the
monodromies, analogous to what we have discussed for the torus, a
couple of first integrals of two $O(2,1)$ sigma models.

Moreover both the first set of equations (\ref{4.11})
coming from the embedding
equations (\ref{3.4}) that the second one (\ref{4.21})
are solved if we add to the system
of conditions (\ref{5.1}) another fundamental equation:

\beq e^{2\phi_f} = e^{2\phi_g} = \frac{\Lambda}{4sin^2 T} e^{2\phi} .
\label{5.2} \eeq

In this way, the static equation of Sine-Gordon type for  $\phi$ is
solved automatically as in the case of the torus, and moreover the
following integral condition holds:

\beq \int_S \nabla^2 \phi = 4 \int_S \frac{df \wedge
d\overline{f}}{1-|f|^2} = 4 \int_S \frac{dg \wedge
d\overline{g}}{1-|g|^2} . \label{5.3} \eeq

\section{A corollary: scattering of black holes }

As a corollary for these set of simplifyed equations we can consider
the scattering of back holes \cite{a34}, in which the gauge choice can
be taken again as the condition $\gamma = 0 $ with the only change
that $T
\rightarrow T + \frac{\pi}{2} $.

One of the problems that were bothering me about building a solution
for the scattering of black holes, was the fact the gauge choice
conformal $g_{zz}=0$ cannot be necessarily global but it can be
extended only until the region external to the horizons, where the
time is well defined and distinct from a spatial coordinate. Therefore
there is not only the spatial end of the universe, but also two
internal horizons on which the condition $|f|=|g|=1$ must hold.

Necessarily we have to impose that the locus of points defined by the
first condition $|f|=1$
coincides with the locus of points defined by the second condition
$|g|=1$. But this requirement of physical consistency is already
inside the equation of consistency of the gauge $\gamma =0$:

\beq \phi_f = \phi_g , \label{6.1} \eeq

because whenever the first condition holds implies that $e^{2\phi_f}$
is divergent as $1/ (1  - {|f|}^2)$, and this equation (\ref{6.1})
automatically requires that there is an analogous divergency of
$e^{2\phi_g}$ as $1/ (1  - {|g|}^2)$. Therefore the set of equations
proposed (\ref{5.1}) and (\ref{5.2}) is complete and must contain the
physically consistent solution for the scattering of black holes.
There are no extra conditions to be added since these are already
contained in the equation (\ref{6.1}).

We can suggest how to solve the monodromy conditions directly, by
requiring that our unknown $f$ is analytic with respect to an
intermediate variable $w$ ( this partial analyticity tell us how to
makes sense to the word cut in a non-analytic setting ),
\beq f = f(w(z,\overline{z},t), \ \ \ \ \ f(w) = {w(z,\overline{z},t)}^{c-1}
\frac{F(a,b,c; w(z,\overline{z},t))}{F(a-c+1,b-c+1,2-c: w(z,\overline{z},t))}
, \label{6.2} \eeq which is single-valued with respect to the physical
complex variable $z$, but it is also a non trivial reparametrization
of it. A quite non trivial constraint on the choice for the
reparametrization $w(z,\overline{z},t)$ comes from the requirement
that the whole solution satisfies the $O(2,1)$ sigma model.

In this way  the problem of defining the geodesic equations for the
particles or the black holes is reduced to find the time-dependent
part of this reparametrization, which at the end is responsible for
their motion. Instead the monodromy conditions are already satisfied
at all orders by a careful choice of the coefficients for the
hypergeometric function. We will show in a forecoming paper what type
of solution is obtained, by doing perturbation with respect to the
parameter $\Lambda$ \cite{a22}.

\section{ Discussion}

We have shown here that the first-order formalism, that has allowed us
to solve the $N$-body problem, can be extended to the case of Riemann
surfaces in $(2+1)$ Gravity with cosmological constant.

We have found how to recover from the solution of two  $O(2,1)$
$\sigma$-model a general solution of all Einstein equations. The
solution is also characterized by an analytic function $N(z)$ which is
the component $K_{zz}$ of the extrinsic curvature tensor. The two
solutions of the $O(2,1)$ sigma model map the $z$-complex plane with
branch cuts into a direct product of unit disks with hyperbolic
metric. The holonomies of $f$ and $g$ are elements of $SU(1,1)\otimes
SU(1,1)$ and isometries of the corresponging hyperbolic metric.
Therefore we can delimitate $Im f \otimes Im g$ into a direct product
of geodesic polygons inside the two unit disk. This property is not
generally true for the analogous mappings $f$and $g$ of the
$N$-particle, since in that case there is a boundary to the spatial
slice of the universe, ad the limit $|f|=1$ and $|g|=1$ are reached.
In the case of the scattering of black holes these limit are reached
not only to represent the spatial infinity but also their horizons.
Instead the peculiar signal of topology should be that these mappings
produce a tessellation of the two unit disks, a property which is not
true for the particle case. The line defined by $|f|=1$ is not reached
by the Riemann surface solutions.

We have given explicit solutions for the mapping functions $f$
and $g$ for the torus and for all Riemann surfaces having $SO(2,1)$
holonomies. It turns out that in both cases the inverse
mapping $z = z(f)$ is a single-valued function, i.e. an automorphic function.
The $N(z)$ function for the torus is related to the quadratic
holomorphic differential, a property that is probably true for
all Riemann surfaces.

We have found that the $SO(2,2)$ holonomies have a quite simple
particle interpretation. They determine the evolution for the
branch points, which move as timelike geodesic of the (\ref{3.8}) hyperboloid
in the embedding flat four-dimensional space-time.

The moduli trajectories, which have to be geodesic of the metric of
Teichmuller space, should be a consequence of the geodesic
trajectories of the branch points.

The next step would be to find a quantization scheme which takes into account
this classical reduction of three dimensional gravity in
two-dimensional field theories.

{\bf Acknowledgements }

I would like to thank A. Cappelli and M. Ciafaloni for useful
discussions.

\appendix

\section{Appendix - First-order formalism in the limit $\Lambda
\rightarrow 0$ }

We would like to clarify how the second-order equations are solved by
the simplified choice of the first-order solution. Let us notice
firstly that the conformal factor can be written in two equivalent
ways:

\begin{eqnarray}
 e^{2\phi} & = & \frac{N\overline{N}}{\partial_z f
\partial_{\overline{z}} \overline{f} } {( 1 - f \overline{f} )}^2 =
\frac{N\overline{N}}{\partial_z g
\partial_{\overline{z}} \overline{g} } {( 1 - g \overline{g} )}^2 =
\nonumber \\
& = & \frac{4 sin^2 T}{\Lambda} \frac{\partial_z \overline{f}
\partial_{\overline{z}} f }{{(1-f\overline{f})}^2} =
\frac{4 sin^2 T}{\Lambda} \frac{\partial_z \overline{g}
\partial_{\overline{z}} g }{{(1-g\overline{g})}^2}
\nonumber \\
I_f & = & \frac{\partial_z f \partial_z
\overline{f}}{{(1-f\overline{f})}^2}
= \frac{\sqrt{\Lambda} e^{-i T}}{2 sin T} N(z) \ \ \ \ \ \
I_g =  \frac{\partial_z g \partial_z
\overline{g}}{{(1-g\overline{g})}^2}
= \frac{\sqrt{\Lambda} e^{i T}}{2 sin T} N(z).
\label{A.1} \end{eqnarray}

The equation for the conformal factor $\phi$ is solved because it can
be written as:

\beq \nabla^2 \phi = 4 \frac{\partial_z f \partial_{\overline{z}}
\overline{f} - \partial_z \overline{f} \partial_{\overline{z}} f }{
{(1- f \overline{f})}^2 } =  4 \frac{\partial_z g \partial_{\overline{z}}
\overline{g} - \partial_z \overline{g} \partial_{\overline{z}} g }{
{(1- g \overline{g})}^2 } . \label{A.2} \eeq In this way it is
pratically identical to the equation of motion that we solved for the
torus case in $(2+1)$gravity ( see Appendix of ref. \cite{a19} ). The
only difference of this case is that the same mechanism of solution is
repeated twice, once for the function $f$ and the other time for the
function $g$.

Analogously the equation for  $\alpha$is solved by looking at the
definition of the field $\alpha$ in terms of first-order quantities:

\beq \partial_z \partial_{\overline{z}} \alpha = \partial_z
\partial_{\overline{z}} V^A \cdot \partial_0 X^a + \partial_z V^A
\cdot \partial_0 \partial_{\overline{z}} X^A + \partial_{\overline{z}} V^A
\cdot \partial_0 \partial_z X^A + V^A \cdot \partial_z
\partial_{\overline{z}} X^A . \label{A.3}\eeq

By rearranging the embedding equations (\ref{3.4}) we can find the
following properties:
\begin{eqnarray} \partial_z\partial_{\overline{z}} V^a & = &
\left( 2 N \overline{N}
e^{-2\phi} + \frac{k^2}{2} e^{2\phi} \right) V^A + \frac{\Lambda}{2}
k(t) e^{2\phi} X^A \nonumber \\
 V^a \cdot \partial_0 \partial_z \partial_{\overline{z}} X^A & = &
\frac{\Lambda}{2} e^{2\phi} \alpha + \partial_0
 \left( \frac{ k(t)}{2}
e^{2\phi} \right) \nonumber \\
\partial_z V^A \cdot \partial_0 \partial_{\overline{z}} X^A & = & k(t)
\partial_z X^A \cdot \partial_0 \partial_{\overline{z}} X^A .
\label{A.4} \end{eqnarray}

By summing up all the contribution the second order equation for
$\alpha$ is naturally solved.

\end{document}